\RequirePackage{fix-cm}

\documentclass[twocolumn,epjc3]{svjour3}  
\RequirePackage{graphicx}
\RequirePackage{amssymb}
\RequirePackage{amsmath,latexsym} 
\RequirePackage{longtable}
\RequirePackage{tabularx}
\RequirePackage{dcolumn}
\RequirePackage{bm}
\RequirePackage{color}
\RequirePackage{hyperref} 

\newcommand{\mev}{\mbox{MeV}}
\newcommand{\gev}{\mbox{GeV}}
\newcommand\degree{{\mathrm o}}

\newcommand\bge{\begin{equation}}
\newcommand\ene{\end{equation}}
\newcommand\bgea{\begin{eqnarray}}
\newcommand\enea{\end{eqnarray}}

\newcommand\ls{\raise 1.5pt\hbox{$\,<\;$}\kern -10.5pt\lower3.5pt
          \hbox{$\sim$}\kern 1.5pt} 
\newcommand\gs{\raise 1.5pt\hbox{$\,>\,$}\kern -9.5pt\lower3.5pt
          \hbox{$\sim$}\kern 1.5pt} 

\newcommand{\eep}{e^+e^-\to e^+e^-\pi^0}
\journalname{Eur. Phys. J. C}

\begin{document}

\title{On the possibility to measure the~$\pi^0\to\gamma\gamma$
  decay width and the~$\gamma^\ast\gamma\to\pi^0$ transition form
  factor with the {KLOE-2}~experiment}

\titlerunning{$\pi^0\to\gamma\gamma$ and $\gamma^\ast\gamma\to\pi^0$
              with KLOE-2}

\author{
D.~Babusci\thanksref{addr1}
\and
H.~Czy\.z\thanksref{addr4}
\and
F.~Gonnella\thanksref{addr2,addr6}
\and
S.~Ivashyn\thanksref{e1,addr5}
\and
M.~Mascolo\thanksref{addr2,addr6}
\and
R.~Messi\thanksref{addr2,addr6}
\and
D.~Moricciani\thanksref{e2,addr6}
\and
A.~Nyffeler\thanksref{addr3}
\and
G.~Venanzoni\thanksref{addr1}
{and~KLOE-2~Collaboration}\thanksref[*]{e3}}

\thankstext{e1}{e-mail: s.ivashyn@gmail.com}
\thankstext{e2}{e-mail: dario.moricciani@roma2.infn.it}
\thankstext[*]{e3}{See Appendix A}

\institute{INFN, Laboratori Nazionali di Frascati, Frascati I-00044, Italy \label{addr1}
\and
Institute of Physics, University of Silesia, Katowice PL-40007, Poland \label{addr4}
\and
Dipartimento di Fisica, Universit\`a ``Tor Vergata'', Roma I-00133 , Italy \label{addr2}
\and
INFN, Sezione Roma ``Tor Vergata'', Roma I-00133, Italy \label{addr6}
\and
A.I. Akhiezer Institute for Theoretical Physics, NSC ``Kharkiv
Institute for Physics and Technology'', Kharkiv UA-61108, Ukraine \label{addr5}
\and
Regional Centre for Accelerator-based Particle Physics,
Harish-Chandra Research Institute,
Chhatnag Road, Jhusi,
Allahabad - 211 019, India \label{addr3}
}

\date{Received: date / Accepted: date}

\maketitle

\begin{abstract}
 A possibility of KLOE-2 experiment to measure
 the width $\Gamma_{\pi^0 \to \gamma\gamma}$
 and the $\pi^0\gamma\gamma^\ast$ form factor $F(Q^2)$
 at low invariant masses of the virtual photon 
 in the space-like region is considered.
 This measurement is an important test of the strong interaction
 dynamics at low energies.
 The feasibility is estimated on the basis
 of a Monte-Carlo simulation. 
 The expected accuracy for $\Gamma_{\pi^0 \to \gamma\gamma}$
 is at a per cent level, which is better than 
 the current experimental world average and theory.
 The form factor will be measured for
 the first time at $Q^2 \leq 0.1~\gev^2$ 
 in the space-like region.
 The impact of these measurements 
 on the accuracy of 
 the pion-exchange contribution to the hadronic light-by-light scattering
 part of the anomalous magnetic moment of the muon is also discussed.
\end{abstract}

\section{Introduction}
\label{sec:intro}
The QCD Green's function $\langle VVA \rangle$ exhibits the 
axial anomaly of Adler, Bell and Jackiw~\cite{Adler:1969gk,Bell:1969ts}
(non-conservation of the axial vector current), which
is responsible for
the decay $\pi^0 \to \gamma\gamma$. The anomaly is a
pure one-loop effect (triangle diagram) and receives corrections
neither perturbatively~\cite{Adler:1969er} nor non-perturbatively~\cite{tHooft:1980xb}. 
It bridges in QCD the strong dynamics of infrared physics at low
energies (pions) with the perturbative description in terms of quarks
and gluons at high energies. The anomaly allows therefore to gain
insights into the strong interaction dynamics of QCD and has received
great attention from theorists over many years. Due to the recent
advances, the decay width $\Gamma_{\pi^0 \to \gamma\gamma}$ is now
predicted with a $1.4\%$ accuracy: $\Gamma^{{\rm theor}}_{\pi^0 \to
  \gamma\gamma} = 8.09 \pm
0.11~\mbox{eV}$~\cite{Kampf:2009tk,Bijnens:2010pa}.
The major experimental information on this decay comes from the
photo-production of pions on a nuclear target via the Primakoff
effect~\cite{Primakoff:1951pj}. 
The most precise value of the pion lifetime cited by PDG~\cite{Nakamura:2010zzi}
comes from a direct decay measurement~\cite{Atherton:1985av}. 
It can be related  to the two-photon width 
via the $\pi^0 \to\gamma\gamma$ branching fraction.
Until recently, the experimental world
average of $\Gamma^{{\rm PDG}}_{\pi^0 \to \gamma\gamma} = 7.74 \pm
0.48$~eV~\cite{Nakamura:2010zzi} was only known to $6.2\%$ precision.
Due to the poor agreement between the existing data,
the PDG error of the width average is inflated (scale factor $2.6$) 
and it gives an additional motivation for new precise measurements.
The PrimEx Collaboration, using a Primakoff effect
experiment at JLab, has achieved $2.8\%$ precision,
reporting the value $\Gamma_{\pi^0 \to
 \gamma\gamma}=7.82\pm0.14\pm0.17$~eV~\cite{Larin:2010kq},
 but this result is not yet included in the PDG average.
There are plans to further reduce the uncertainty to the per cent level.

Though theory and experiment are in a fair agreement, a better
experimental precision is needed to really test the theory predictions.
The Primakoff effect-based experiments suffer from model dependence due to
the contamination by the coherent and incoherent conversions in the
strong field of a nucleus~\cite{Kaskulov:2011ab:preprint}.
Therefore, a measurement using a completely different method
(Section~\ref{sec:width_extraction}), which can reach a similar
accuracy, is highly desirable.  

The first aim of this letter is to demonstrate that 
a per cent level of precision can be achieved in the
measurement of $\Gamma_{\pi^0 \to \gamma\gamma}$ by the
KLOE-2 experiment at Frascati
(Section~\ref{sec:simulation:width}), where
the first phase of data taking (step-$0$)
is expected to have the integrated luminosity of $5$~fb$^{-1}$~\cite{AmelinoCamelia:2010me}.

Putting the pion on-shell
in the QCD Green's function $\langle VVA \rangle$, 
one can defines the $\pi^0\gamma^*\gamma^*$ form factor 
${\cal F}_{\pi^0\gamma^*\gamma^*}(q_1^2, q_2^2)$    
\bgea
&& i \int d^4 x  e^{i q_1 \cdot x}  \langle 0 | T \{ j_\mu(x) j_\nu(0) \}|
\pi^0(q_1 + q_2) \rangle \nonumber \\ 
& & \quad\quad = \varepsilon_{\mu\nu\rho\sigma} q_1^\rho q_2^\sigma 
 {\cal F}_{\pi^0\gamma^\ast\gamma^\ast}(q_1^2, q_2^2),  
  \label{eq:def_FF}
\enea
where $j_\mu$ is the electromagnetic current of the light quarks
($u,d,s$), $\varepsilon_{\mu\nu\rho\sigma}$ is the Levi-Civita symbol
 and $q_1$ and $q_2$ are the 4-momenta of the off-shell
photons. 
The form factor for
real photons is related to the $\pi^0\to\gamma\gamma$
decay width:
\bgea
\label{eq:ff-normalization}
{\cal F}^2_{\pi^0\gamma^\ast\gamma^\ast}(q_1^2=0, q_2^2=0) 
 &=& 
 \frac{4}{\pi \alpha^2 m_\pi^3}
 \Gamma_{\pi^0 \to \gamma\gamma}.
\enea

The form factor ${\cal F}_{\pi^0\gamma^*\gamma^*}(q_1^2, q_2^2)$ as a
function of both photon virtualities 
has never been studied experimentally in the space-like region,
and studied with very limited accuracy
in the time-like region~\cite{Abouzaid:2008cd}. 
The pion-photon transition form factor $F(Q^2)$ 
with one on-shell and one off-shell photon 
\bge
F(Q^2) \equiv {\cal F}_{\pi^0\gamma^\ast\gamma^\ast}(-Q^2, q_2^2=0),
\quad\quad Q^2 \equiv - q^2
\ene
has been measured in the experiments CELLO~\cite{Behrend:1990sr},
CLEO~\cite{Gronberg:1997fj} 
and BaBar~\cite{Aubert:2009mc} at 
large space-like momenta ($Q^2 \geq 0.5~\gev^2$).

The second aim of this letter is to show that the KLOE-2 
experiment can perform the first measurement  
of $F(Q^2)$ in the space-like region
in the vicinity of the origin, namely for $0.01 < Q^2 < 0.1~\mbox{GeV}^2$
(Sections~\ref{sec:width_extraction},~\ref{sec:simulation:ff}).

The third aim of this letter is to estimate the impact of 
the proposed KLOE-2 measurements 
on the evaluation of the Standard Model prediction 
for the anomalous magnetic moment of the muon, $a_\mu$
(Section~\ref{sec:discussion:impact}).
The theoretical value of $a_\mu$ is currently limited by uncertainties from the hadronic
vacuum polarization and the hadronic light-by-light (LbyL) scattering
contribution. 
The value of the latter is currently obtained using hadronic
models (e.g.,~\cite{Melnikov:2003xd,Prades:2009tw,Nyffeler:2009tw,Jegerlehner:2009ry})
and leads to an uncertainty in $a_\mu$ of 
$(26 - 40) \times 10^{-11}$
   (the Dyson- Schwinger approach~\cite{Goecke:2010if} is still far
    from this value),
which is almost as large as the one from hadronic
vacuum polarization $\sim (40-50) \times
10^{-11}$~\cite{Davier:2010nc,Hagiwara:2011af}. 
For comparison, the precision of the 
Brookhaven $g-2$ experiment is $63 \times 10^{-11}$~\cite{Bennett:2006fi}. 
In view of the proposed new $g-2$
experiments at Fermilab~\cite{Carey:2009zzb} and JPARC~\cite{JParc:amm:2010}
with a precision of $15 \times 10^{-11}$, the hadronic LbyL
contribution needs to be controlled much better, in order to fully profit from
these new experiments to test the Standard Model and constrain New
Physics.  
According to model calculations, the exchange of neutral pions
yields the numerically dominant contribution, $a_\mu^{\mathrm{LbyL};\pi^0}$, 
 to the final result for hadronic LbyL scattering. 

The conclusions are given in Section~\ref{sec:conclusions}.


\section{The basics of width and form factor measurement}
\label{sec:width_extraction}
Since the original proposal of Low~\cite{Low:1960wv}
to measure the width of a neutral pion decay into two photons
using the $e^+e^- \to e^+e^-\pi^0$ process, only at DESY 
this measurement has been done using this method,
with the result
 $\Gamma_{\pi^0 \to \gamma\gamma}=7.7\pm0.5\pm0.5$~eV~\cite{Williams:1988sg}. 
It was stressed in~\cite{Parisi:1972iv}
and~\cite{Terazawa:1973tb} that
for a precision measurement of $\pi^0$ width 
via the ``$\gamma\gamma$ fusion'' (${\gamma\gamma \to  \pi^0}$) process, 
one needs to improve the original Low's proposal.
Namely, instead of a no-tag experiment (like~\cite{Williams:1988sg}), 
one should perform a lepton double-tagging at small angles.
For the $\phi$--factory DA$\Phi$NE
a detailed study was performed in~\cite{Alexander:1993rz},
where a lepton tagging system was proposed,
which, however, was not installed.
This proposal was reconsidered for the KLOE-2~\cite{AmelinoCamelia:2010me}
experiment at DA$\Phi$NE.
The two Low Energy Taggers (LET)~\cite{Babusci:2009sg} and 
two High Energy Taggers (HET)~\cite{Archilli:2010zza}
were specially designed for this experiment
and will allow detection of electrons and positrons, scattered
at very small polar angles ($\theta < \theta_{max} \approx 1^\degree$)
in two domains of lepton energy.
For the present letter only information coming from HET 
is considered. 
The effect of the LET detectors and of those to be installed in the 
near future (see Ref.~\cite{AmelinoCamelia:2010me}) will be 
the subject of a forthcoming investigation.

One can extract the value of the partial decay width from data, using the formula
\bgea
\label{eq:DESY-extr}
\Gamma_{\pi^0 \to \gamma\gamma} &=&
 \frac{N_{\pi^0}}{\varepsilon \;\mathcal{L}}
 \;
 \frac{\tilde{\Gamma}_{\pi^0 \to \gamma\gamma}}{\tilde{\sigma}_{e^+e^- \to e^+e^-\pi^0}}
,
\enea
where $N_{\pi^0}$ is the number of detected pions, 
$\varepsilon$ accounts for the detection acceptance  and efficiency,
$\mathcal{L}$ is the integrated luminosity,
$\tilde{\Gamma}_{\pi^0 \to \gamma\gamma}$ is the model $\pi^0$ width 
and 
$\tilde{\sigma}_{e^+e^- \to e^+e^-\pi^0}$ is the cross section obtained
with a Monte Carlo simulation using the same model as for the 
$\tilde{\Gamma}_{\pi^0 \to \gamma\gamma}$ calculation.

The form factor $F(Q^2)$ 
can be evaluated through the relation 
\begin{equation}
\label{ff}
 \frac{F^2(Q^2)}{F^2(Q^2)_{MC}} = \frac{(\frac{d\sigma}{dQ^2})_{data}}{(\frac{d\sigma}{dQ^2})_{MC}}
,
\end{equation}
where $(\frac{d\sigma}{dQ^2})_{data}$ is the experimental  
differential cross section, and  
$(\frac{d\sigma}{dQ^2})_{MC}$ is the Monte Carlo one obtained with  
the form factor ${F(Q^2)_{MC}}$.   

\section{Feasibility of the $\pi^0$ width measurement}
\label{sec:simulation:width}
The $\pi^0$ production in the process $\eep$ is simulated 
with EKHARA~\cite{Czyz:2010sp} Monte Carlo event generator. 
The simulated signal is given by 
the t-channel amplitude ($\gamma^*\gamma^*\to \pi^0$).

Since a stand-alone  EKHARA version 
works in the CM (center of mass) frame of incident leptons 
and does not simulate the pion decays, 
it has been modified to take into account the DA$\Phi$NE crossing 
angle between the  incoming beams ($\theta_{e^+e^-} \approx 51.3$~mrad)
and the decay of the $\pi^0$  into two photons.

In order to simulate the DA$\Phi$NE optics, EKHARA has been interfaced 
with the BDSIM package~\cite{Agapov:2009zz} which allows to trace 
the emitted electron (positron) through the magnetic elements 
of DA$\Phi$NE layout.
In this way only few of them ($\sim 2\%$) reach the HET detectors, 
providing a realistic estimate of its acceptance.
In the following the coincidence of the HET detectors will be required, 
which selects the energy of the final leptons to be between 420 and 460 MeV.

\begin{figure}[htbp]
 \resizebox{0.48\textwidth}{!}{
  \includegraphics{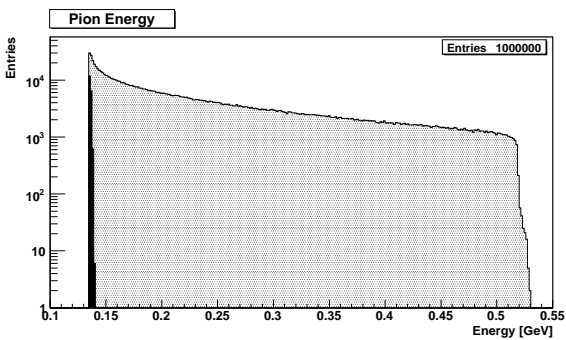} 
   }
\caption{The $\pi^0$ energy (in the laboratory frame)
         distribution with (dark) and 
         without (light-gray)  HET-HET coincidence.
\label{fig:matteo1}
    }
\end{figure}

\begin{figure}[htbp]
 \resizebox{0.48\textwidth}{!}{
  \includegraphics{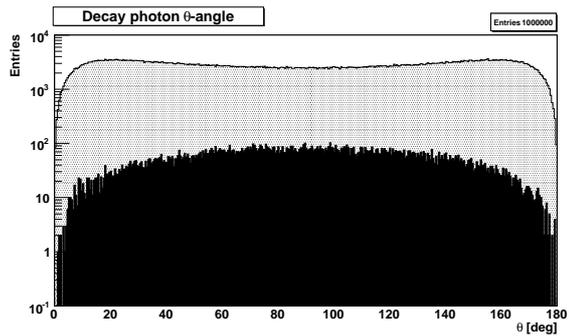} 
   }
\caption{Polar angle (in the laboratory frame)
         distribution of decay photons 
         from $\pi^0$ with (dark) and without (light-dark) 
         the HET-HET coincidence.
\label{fig:matteo2}
    }
\end{figure}

Figure~\ref{fig:matteo1} shows the energy of the emitted $\pi^0$ 
in the $\gamma\gamma$ process: as can be seen, the request of the 
 HET-HET coincidence allows us to select $\pi^0$ almost at rest 
 (dark region), compared with the no-tag case (light-gray). 
 Since the $\pi^0$ decays almost at rest, 
 most of the photons from its decay are emitted with large polar 
angle (defined as the angle between the direction 
               of the photon and the beam axis), 
as shown in Fig.~\ref{fig:matteo2}. 
In particular, about 95\% of the photons are emitted above $25^\circ$ 
(and below $155^\circ$), resulting in a large acceptance for photons 
reaching the KLOE Electromagnetic Calorimeter (EMC)~\cite{Adinolfi:2002zx}.

By requiring both photons in the barrel of the EMC
(between $50^\circ$ and $130^\circ$) 
and the HET-HET coincidence, a value for the acceptance 
$\epsilon_{acc}$ of 1.2\% is obtained.
Since the total cross-section of $\eep$ at $\sqrt{s} = 1020$~MeV 
is $\sigma_{tot} \approx 0.28$~nb, a cross-section of about $3.4$~pb  
is obtained within the acceptance cuts.
The integrated luminosity $\mathcal{L}$ at DA$\Phi$NE
required to reach a $1~\%$ statistical error is:
\bgea 
 \mathcal{L} &=& \frac{10000}{\sigma_{tot} \,\epsilon_{acc}\, \epsilon_{det}}
 \approx \frac{3}{\epsilon_{det}} \,\, \text{fb}^{-1},
\enea
where the efficiency $\epsilon_{det}$
due to trigger, reconstruction and analysis criteria
is estimated to be about $50~\%$.
Therefore, the required data sample 
can be obtained during the first phase 
(about one year) of data taking.

\begin{figure}[htbp]
 \resizebox{0.48\textwidth}{!}{
  \includegraphics{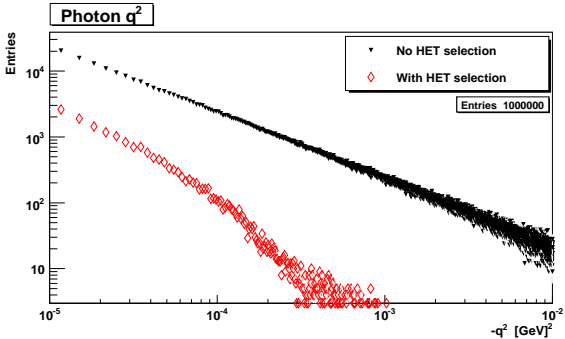} 
   }
\caption{\label{Fig.4.10} 
          Distribution of the photon virtuality 
          in $\gamma^*\gamma^*\to\pi^0$. 
          The lepton double tagging (HET-HET)
          selects the events (red diamonds) with 
          small virtuality of the photons.
         }
\end{figure}

Extraction of the width $\Gamma(\pi^0 \rightarrow \gamma\gamma)$ 
with $\sim 1\%$ accuracy requires  
a very good control of the systematic errors.
From the experimental side, the clean signature of the process, 
the use of the KLOE detector and the  HET-HET coincidence 
should allow to keep the systematic effects under control at 
the required level of precision.
A possible background to this measurement comes from the double
radiative Bhabha scattering, which has the same signature as our signal.
An extensive simulation of this background (based on $10^8$ events
generated with Babayaga MC~\cite{CarloniCalame:2000pz,Balossini:2006wc})
shows that no events survive
the coincidence of HET for electrons and positrons and the KLOE acceptance
for photon, and therefore the expected contribution is negligible.
From the theoretical side, the systematic errors can arise also from 
accuracy of the generator, e.g., due to missing radiative corrections
and the uncertainty in the modeled 
$\pi^0\gamma^\ast\gamma^\ast$  transition form factor. 
In order to reduce the former effect, 
the radiative corrections are planned to be introduced
in the EKHARA generator.
For the latter effect, a numerical simulation with different
formulae for the form factor can be performed.
The HET-HET coincidence, imposed in such a simulation,
leads to a significant restriction
on the photon virtuality in $\gamma^*\gamma^*\to\pi^0$:
for most of the events one has $|q^2| < 10^{-4}$ GeV$^2$,
as shown in Fig.~\ref{Fig.4.10}.
Thus, for the KLOE-2 case the possible effect of the photon virtualities
which can influence the accuracy of eq.~(\ref{eq:DESY-extr}) is
negligible.
Our simulation shows that the uncertainty
in the measurement of $\Gamma(\pi^0 \rightarrow \gamma\gamma)$
due to the form factor parametrization in the generator is expected to be
less than $0.1~\%$.
 
\section{Feasibility of the $\gamma^\ast \gamma \pi^0$ 
transition form factor measurement}
\label{sec:simulation:ff} 
By requiring one lepton 
inside the KLOE detector 
($20^{\circ} < \theta < 160^{\circ}$, corresponding to
$0.01 < |q_1^2| < 0.1  \, \text{GeV}^{2}$) 
and the other lepton in the HET detector 
(corresponding to $|q_2^2| \lesssim 10^{-4} \, \text{GeV}^{2}$
 for most of the events) 
one can measure the differential cross section 
$({d\sigma}/{dQ^2})_{data}$, where $Q^2\equiv-q_1^2$. 
Using eq.~(\ref{ff}), the form factor $|F(Q^2)|$
can be extracted from this cross section. 

\begin{figure}[htbp]
 \resizebox{0.48\textwidth}{!}{
  \includegraphics{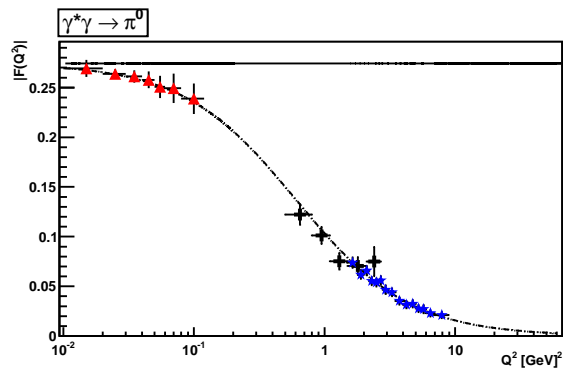} 
   }
\caption{
     Simulation of KLOE-2 measurement of
     $F(Q^2)$  
     (red triangles) with statistical errors for $5$~fb$^{-1}$. 
     Dashed line is the $F(Q^2)$ form factor according to  
     LMD+V model~\cite{Knecht:2001xc},  
     solid line is $F(0)$ given by Wess-Zumino-Witten term, eq.~(\ref{eq:WZW:ff:norm}).
     CELLO~\cite{Behrend:1990sr} (black crosses) and CLEO~\cite{Gronberg:1997fj} 
     (blue stars) data 
     at high $Q^2$ are also shown for illustration.
    \label{contri}
    }
\end{figure}

The simulation has been performed using 
a lowest meson dominance ansatz with two vector multiplets (LMD+V)
for the form factor ${\cal F}_{{\pi^0}\gamma^*\gamma^*}$,
which is available in EKHARA.
The LMD+V ansatz is based on large-$N_C$ QCD matched 
to short-distance constraints from 
the operator-product expansion (OPE), see the Ref.~\cite{Knecht:2001xc}.
In the following we use the definition of the LMD+V parameters
$\bar{h}_5=h_5+h_3m_\pi^2$ and $\bar{h}_7=h_7+h_6m_\pi^2+h_4m_\pi^4$.
Figure~\ref{contri} shows the expected experimental uncertainty 
(statistical) on $F(Q^2)$ achievable 
at KLOE-2 with an integrated luminosity of $5$~fb$^{-1}$.
In this measurement the detection efficiency is different
and is estimated to be about~$20\%$.
From our simulation we conclude that a statistical uncertainty of less than
 $6\%$ for every bin is feasible. 

Having measured the form factor, one can evaluate also
the slope parameter $a$ of the form factor at the origin\footnote{
      We would like to stress that 
      the $q^2$ range of KLOE-2 measurement is not small enough
      to use the linear approximation
      ${\cal F}_{\pi^0\gamma^\ast\gamma^\ast}(q^2, 0) = {\cal F}_{\pi^0\gamma^\ast\gamma^\ast}(0, 0)
       (1 + q^2\, a/{m_\pi^2})$
      because the higher order terms are not negligible.
     }
\begin{equation}
\label{eq:slope:1}
 a \equiv m_\pi^2 \frac{1}{{\cal F}_{\pi^0\gamma^\ast\gamma^\ast}(0, 0) } 
          \left.\frac{d\, {\cal F}_{\pi^0\gamma^\ast\gamma^\ast}(q^2, 0) }
 {d\, q^2}\right|_{q^2=0}.
\end{equation}
Though for {\it time-like} photon
virtualities ($q^2 > 0$), the slope can be measured directly 
in the rare decay $\pi^0 \to e^+ e^-\gamma$, 
the current experimental uncertainty is very big~\cite{Farzanpay:1992pz,MeijerDrees:1992qb}.
The PDG average value of the slope parameter 
is quite precise, $a = 0.032 \pm 0.004$~\cite{Nakamura:2010zzi},
and it is dominated by the CELLO result~\cite{Behrend:1990sr}.
In the latter, a simple vector-meson dominance (VMD)
form factor parametrization was fitted to 
the data~\cite{Behrend:1990sr} and then the slope
was calculated according to eq.~(\ref{eq:slope:1}).
Thus the CELLO procedure for the slope calculation 
suffers from model dependence not accounted for
in the error estimation.
The validity of such a procedure has never been verified,
because there were no data at $Q^2 < 0.5~\gev^2$.
Therefore, filling of this gap in $Q^2$
by the KLOE-2 experiment can 
provide a valuable test of the form factor parametrizations.

When the normalization of the form factor is fixed to the
decay width $\pi^0 \to \gamma\gamma$ or to some effective pion decay
constant $F_\pi$,
the VMD and (on-shell) LMD+V models have only one free 
parameter\footnote{In the Brodsky-Lepage
  ansatz~\cite{Lepage:1979zb,Lepage:1980fj,Brodsky:1981rp} 
  the parameter $F_\pi$ fixes the normalization {\em and}
  the asymptotic behavior at the same time. 
  Comparison with data from
  CELLO and CLEO shows that the asymptotic behavior is off by about
  20\%, once the normalization is fixed from $\pi^0 \to
  \gamma\gamma$.}.
For VMD this parameter is the vector-meson mass $M_V$
(sometimes denoted by $\Lambda_{\pi^0}$) and for LMD+V this is $\bar{h}_5$, 
once we put $h_1 = 0$ to get the $1/Q^2$ behavior
for large $Q^2$, as expected from theoretical
arguments~\cite{Lepage:1979zb,Lepage:1980fj,Brodsky:1981rp}. 
It is a priori not clear why only one
parameter should be sufficient to describe the behavior of the form
factor simultaneously at low momenta (slope at the origin) and at
large momenta (asymptotic behavior, related to perturbative QCD / OPE
near the light-cone).
Since the available data~\cite{Behrend:1990sr,Gronberg:1997fj,Aubert:2009mc}
cover only the relatively high $Q^2 > 0.5~\gev^2$ region,
a new measurement by KLOE-2 at $Q^2 < 0.1~\gev^2$ would help to
verify the consistency of the parametrizations of
the form factor $F(Q^2)$.

\section{Impact 
  on the hadronic light-by-light
  scattering contribution to the muon $g-2$} 
\label{sec:discussion:impact}

The value of the pion-exchange part $a_\mu^{\mathrm{LbyL};\pi^0}$
of the hadronic LbyL contribution to $a_\mu$
is currently obtained using hadronic models and
any experimental information on the transition form factor 
is important in order to constrain the models. 
However, having a good description for
the transition form factor is only necessary, not sufficient,
in order to uniquely determine $a_\mu^{\mathrm{LbyL};\pi^0}$. 

As pointed out in
Refs.~\cite{Jegerlehner:2007xe,Jegerlehner:2008zza}, what enters in
the calculation of the pion-exchange contribution $a_\mu^{{\rm LbyL};
  \pi^0}$ is the fully off-shell form factor ${\cal
  F}_{{\pi^0}^*\gamma^*\gamma^*}((q_1 + q_2)^2, q_1^2, q_2^2)$ (vertex function), where
also the pion is off-shell with 4-momentum $q_1 + q_2$. 
The form
factor defined in eq.~(\ref{eq:def_FF}) with on-shell pions is then
given by ${\cal F}_{\pi^0\gamma^*\gamma^*}(q_1^2, q_2^2) \equiv {\cal
  F}_{{\pi^0}^*\gamma^*\gamma^*}(m_\pi^2, q_1^2, q_2^2)$.
A measurement of the transition form factor 
${\cal F}_{{\pi^0}^*\gamma^*\gamma^*}(m_{\pi}^2, q^2, 0)$ can
only be sensitive to a subset of the model parameters and
in general does not allow to reconstruct the full off-shell form factor.
Therefore, within any given approach, 
the uncertainty of the calculated $a_\mu^{{\rm LbyL};
  \pi^0}$ related to the off-shell pion can
be different and the complete error on $a_\mu^{{\rm LbyL};
  \pi^0}$ should take into account this model dependence.

For instance, the estimate in the LMD+V model $a_{\mu; {\rm
    LMD+V}}^{{\rm LbyL}; \pi^0} = (72 \pm 12) \times 10^{-11}$ given
in Ref.~\cite{Nyffeler:2009tw} is based on the variation of all model
parameters, where $\bar{h}_5 = (6.93 \pm 0.26)~\gev^4$ has been used,
which was obtained in Ref.~\cite{Knecht:2001xc} from a fit to the CLEO
data for the transition form factor $F(Q^2)$. The variation of $\pm
0.26~\gev^4$ in $\bar{h}_5$ only leads to a variation in $a_{\mu; {\rm
    LMD+V}}^{{\rm LbyL}; \pi^0}$ of $\pm 0.6 \times 10^{-11}$.  Within
the off-shell LMD+V model the variation of the parameters related to
the off-shellness of the pion completely dominate the total
uncertainty and will {\it not} be shown in Table~\ref{tab:amu} below.

 \begin{table*}
 \caption{\label{tab:amu} Estimate of KLOE-2 impact
          on the accuracy of $a_\mu^{{\rm LbyL}; \pi^0}$
          in case of one year of data taking ($5$~fb$^{-1}$).
          For calculation we
          used the Jegerlehner-Nyffeler (JN)~\cite{Nyffeler:2009tw,Jegerlehner:2009ry} and  
          Melnikov-Vainshtein (MV)~\cite{Melnikov:2003xd} approaches.
          The values marked with asterisk (*)
          do not contain additional uncertainties coming from
          the ``off-shellness'' of the pion (see the text).
          Data sets used for fits (A0, A1, A2, B0, B1, B2) --- see the text, eq.~(\ref{eq:fitdatasets}). }
 \begin{tabularx}{\textwidth}{clcllll}
 Model&Data& $\chi^2/d.o.f.$ &  & Parameters && $a_\mu^{{\rm LbyL}; \pi^0} \times 10^{11}$\\
 \hline
 VMD  & A0 & $6.6/19$
     & $M_V = 0.778(18)$~GeV & $F_\pi = 0.0924(28)$~GeV  && $(57.2 \pm 4.0)_{JN}$\\
 VMD  & A1 & $6.6/19$
     & $M_V = 0.776(13)$~GeV & $F_\pi = 0.0919(13)$~GeV  && $(57.7 \pm 2.1)_{JN}$\\
 VMD  & A2 & $7.5/27$
     & $M_V = 0.778(11)$~GeV & $F_\pi = 0.0923(4)$~GeV   && $(57.3 \pm 1.1)_{JN}$\\
 VMD  & B0 & $77/36$
     & $M_V = 0.829(16)$~GeV & $F_\pi = 0.0958(29)$~GeV  && --- \\
 VMD  & B1 & $78/36$
     & $M_V = 0.813(8)$~GeV & $F_\pi = 0.0925(13)$~GeV  && --- \\
 VMD  & B2 & $79/44$
     & $M_V = 0.813(5)$~GeV & $F_\pi = 0.0925(4)$~GeV   && --- \\
 \hline
 LMD+V, $h_1 = 0$  & A0 & $6.5/19$ 
     &  $\bar{h}_5 = 6.99(32)$~GeV$^4$ & $\bar{h}_7 = -14.81(45)$~GeV$^6$ && $(72.3 \pm 3.5)_{JN}^*$\\
 &  &   &                        &                            && $(79.8 \pm 4.2)_{MV}$\\
 LMD+V, $h_1 = 0$  & A1 & $6.6/19$ 
     &  $\bar{h}_5 = 6.96(29)$~GeV$^4$ & $\bar{h}_7 = -14.90(21)$~GeV$^6$ && $(73.0 \pm 1.7)_{JN}^*$\\
 &  &   &                        &                            && $(80.5 \pm 2.0)_{MV}$\\
 LMD+V, $h_1 = 0$  & A2 & $7.5/27$
     &  $\bar{h}_5 = 6.99(28)$~GeV$^4$ & $\bar{h}_7 = -14.83(7)$~GeV$^6$ && $(72.5 \pm 0.8)_{JN}^*$\\
 &  &   &                        &                           && $(80.0 \pm 0.8)_{MV}$\\
 LMD+V, $h_1 = 0$  & B0 & $65/36$ 
     &  $\bar{h}_5 = 7.94(13)$~GeV$^4$ & $\bar{h}_7 = -13.95(42)$~GeV$^6$ && --- \\
 LMD+V, $h_1 = 0$  & B1 & $69/36$ 
     &  $\bar{h}_5 = 7.81(11)$~GeV$^4$ & $\bar{h}_7 = -14.70(20)$~GeV$^6$ && --- \\
 LMD+V, $h_1 = 0$  & B2 & $70/44$
     &  $\bar{h}_5 = 7.79(10)$~GeV$^4$ & $\bar{h}_7 = -14.81(7)$~GeV$^6$  && --- \\
 \hline
 LMD+V, $h_1 \neq 0$  & A0 & $6.5/18$ 
     &  $\bar{h}_5 = 6.90(71)$~GeV$^4$ & $\bar{h}_7 = -14.83(46)$~GeV$^6$& 
        $h_1 = -0.03(18)$~GeV$^2$ & $(72.4 \pm 3.8)_{JN}^*$\\
 LMD+V, $h_1 \neq 0$  & A1 & $6.5/18$ 
     &  $\bar{h}_5 = 6.85(67)$~GeV$^4$ & $\bar{h}_7 = -14.91(21)$~GeV$^6$& 
        $h_1 = -0.03(17)$~GeV$^2$ & $(72.9 \pm 2.1)_{JN}^*$\\
 LMD+V, $h_1 \neq 0$  & A2 & $7.5/26$ 
     &  $\bar{h}_5 = 6.90(64)$~GeV$^4$ & $\bar{h}_7 = -14.84(7)$~GeV$^6$ &
        $h_1 = -0.02(17)$~GeV$^2$ & $(72.4 \pm 1.5)_{JN}^*$\\
 LMD+V, $h_1 \neq 0$  & B0 & $18/35$ 
     &  $\bar{h}_5 = 6.46(24)$~GeV$^4$ & $\bar{h}_7 = -14.86(44)$~GeV$^6$ &
        $h_1 = -0.17(2)$~GeV$^2$ & $(71.9 \pm 3.4)_{JN}^*$\\
 LMD+V, $h_1 \neq 0$  & B1 & $18/35$ 
     &  $\bar{h}_5 = 6.44(22)$~GeV$^4$ & $\bar{h}_7 = -14.92(21)$~GeV$^6$ &
        $h_1 = -0.17(2)$~GeV$^2$ & $(72.4 \pm 1.6)_{JN}^*$\\
 LMD+V, $h_1 \neq 0$  & B2 & $19/43$ 
     &  $\bar{h}_5 = 6.47(21)$~GeV$^4$ & $\bar{h}_7 = -14.84(7)$~GeV$^6$ &
        $h_1 = -0.17(2)$~GeV$^2$ & $(71.8 \pm 0.7)_{JN}^*$
 \end{tabularx}
 \end{table*}

In contrast to the off-shell LMD+V model, many models
 do not have these additional sources of uncertainty
 (the VMD model, the ans\"atze for the transition form factor used in
 Ref.~\cite{Cappiello:2010uy}, etc.).
Therefore, the precision of the KLOE-2 measurement
can dominate the total accuracy of $a_\mu^{\mathrm{LbyL};\pi^0}$ in such models.

We would like to stress that a realistic calculation 
of $a_\mu^{{\rm LbyL}; \pi^0}$ is {\it not} the purpose
of this letter.
The estimates given below are performed to demonstrate,
within several approaches, an improvement of uncertainty,
which will be possible when the KLOE-2 data appear.
Discussion of the validity of these approaches as
well as the form factor modeling is beyond the scope of
this letter.

As pointed out in Ref.~\cite{Nyffeler:2009uw}, essentially
all evaluations of the pion-exchange contribution (or the pion-pole
contribution with on-shell form factors) use 
the following normalization for the form factor
\bgea
\label{eq:WZW:ff:norm}
{\cal F}_{{\pi^0}^*\gamma^*\gamma^*}(m_\pi^2, 0, 0) 
 &=& 1 / (4 \pi^2 F_\pi)
\enea
derived from the Wess-Zumino-Witten (WZW)
term~\cite{Wess:1971yu,Witten:1983tw}
and the value $F_\pi = 92.4~\mev$ is used 
without any error attached to it\footnote{Note
  that this value of $F_\pi$ is close to $F_\pi = (92.2 \pm
  0.14)~\mbox{MeV}$, as derived from $\pi^+ \to \mu^+ \nu_\mu(\gamma)$
  with 0.15\% precision~\cite{Nakamura:2010zzi}. It leads, 
  according to eqs.~(\ref{eq:WZW:ff:norm}) and~(\ref{eq:ff-normalization}), to
  $\Gamma(\pi^0 \to \gamma\gamma) = 7.73~\mbox{eV}$, which is
  consistent with the current PDG average.
  For a detailed discussion of $F_\pi$ and $\Gamma(\pi^0 \to \gamma\gamma)$
  see, e.g.,~\cite{Kampf:2009tk,Bijnens:2010pa,Li:2011kk:preprint}.}.    
Instead, if one uses the decay width $\Gamma_{\pi^0 \to \gamma\gamma}$ 
for the normalization of the form factor, see
eq.~(\ref{eq:ff-normalization}), an additional source of
uncertainty enters.  
This uncertainty has not been taken into account so far, 
except in the very recent paper~\cite{Dorokhov:2011zf}\footnote{
  In~\cite{Dorokhov:2011zf} the experimental uncertainty
  of the decay width on $a_\mu^{{\rm LbyL}; \pi^0}$ has been estimated
  within the context of the nonlocal chiral quark model.
  }.
In our calculations we account for this normalization issue, using in
the fit:
\begin{itemize}
\item $\Gamma^{{\rm PDG}}_{\pi^0 \to \gamma\gamma} = 7.74 \pm
0.48$~eV~\cite{Nakamura:2010zzi} for a pre-PrimEx case,
\item $\Gamma^{{\rm PrimEx}}_{\pi^0 \to \gamma\gamma} = 7.82 \pm
0.22$~eV~\cite{Larin:2010kq} for a pre-KLOE-2 case,
\item $\Gamma^{{\rm KLOE-2}}_{\pi^0 \to \gamma\gamma} = 7.73 \pm
0.08$~eV for a KLOE-2 simulation (assuming a $1\%$ precision, 
see Section~\ref{sec:simulation:width}).
\end{itemize}

In this Section we assume that the KLOE-2 measurement will be consistent with the
LMD+V and VMD models.  This allows us to use the simulation
(Sections~\ref{sec:simulation:width} and~\ref{sec:simulation:ff}) as
new ``data'' and evaluate the impact of such ``data'' on the precision
of the $a_\mu^{{\rm LbyL}; \pi^0}$ calculation.  In order to do that,
we fit the LMD+V and VMD models to the following data sets:
\bgea
\label{eq:fitdatasets}
 \begin{array}{ll}
   A0: & \mbox{CELLO, CLEO, PDG;} \\
   A1: & \mbox{CELLO, CLEO, PrimEx;} \\
   A2: & \mbox{CELLO, CLEO, PrimEx, KLOE-2;} \\
   B0: & \mbox{CELLO, CLEO, BaBar, PDG;} \\
   B1: & \mbox{CELLO, CLEO, BaBar, PrimEx;} \\
   B2: & \mbox{CELLO, CLEO, BaBar, PrimEx, KLOE-2;}
 \end{array}
\enea
and evaluate $a_\mu^{{\rm LbyL}; \pi^0}$.

Some comments about the BaBar data~\cite{Aubert:2009mc} are in order
here. This measurement of the pion transition form factor does not
show the $1/Q^2$ behavior as expected from earlier theoretical
considerations~\cite{Lepage:1979zb,Lepage:1980fj,Brodsky:1981rp} and
as seen in the CELLO and CLEO data (although in the latter experiments the
$Q^2$ was maybe not yet large enough). The situation is puzzling,
since BaBar observes for the $\eta$ that $Q^2 F(Q^2)$ rises about
three times slower than for the pion, whereas the transition form
factor of the $\eta^\prime$ shows a $1/Q^2$ fall-off, see
Ref.~\cite{BABAR_eta_etaprime}.
Though several approaches exist, which claim to be able to reconcile
the data of Refs.~\cite{Aubert:2009mc} and~\cite{BABAR_eta_etaprime}
(see, e.g., the results of Ref.~\cite{Kroll:2010bf}), 
the strong obstacles for the theory to confront the data~\cite{Aubert:2009mc}
are being widely discussed lately 
(see~\cite{Brodsky:2011xx,Bakulev:2011rp}).

The VMD model always shows a $1/Q^2$ fall-off and therefore is not
compatible with the BaBar data. The LMD+V model has another parameter,
$h_1$, which determines the behavior of the transition form factor for
large $Q^2$. To get the $1/Q^2$ behavior according to
Brodsky-Lepage~\cite{Lepage:1979zb,Lepage:1980fj,Brodsky:1981rp}, one
needs to set $h_1 = 0$. However, one can simply leave $h_1$ as a free
parameter and fit it to the BaBar data, yielding $h_1 \neq
0$~\cite{Nyffeler:2009uw}. In this case the form factor does not
vanish at $Q^2 \to \infty$. Since VMD and LMD+V with $h_1 = 0$ are not
compatible with the BaBar data (as can be seen from the large $\chi^2$
per degree of freedom of the fits of the data sets B0, B1 and B2
below), we will not evaluate $a_\mu^{{\rm LbyL}; \pi^0}$ for these
cases.

For illustration, we use the following two approaches to calculate
 $a_\mu^{{\rm LbyL}; \pi^0}$: 
\begin{itemize}
 \item Jegerlehner-Nyffeler (JN)
       approach~\cite{Nyffeler:2009tw,Jegerlehner:2009ry} 
       with the off-shell pion form factor;
 \item Melnikov-Vainshtein (MV) approach~\cite{Melnikov:2003xd},
       where one uses the on-shell pion form factor in one vertex
       and the other vertex is constant (WZW).
\end{itemize}

Table~\ref{tab:amu} shows the impact of the (existing) PrimEx and the
(future) KLOE-2 measurements on the model parameters (e.g., the
normalization of the form factor) and, consequently, on the
$a_\mu^{{\rm LbyL}; \pi^0}$ uncertainty.  The errors of the fitted
parameters are the {\tt MINOS} ({\tt MINUIT} from {\tt CERNLIB})
parabolic errors. The other parameters of the (on-shell and off-shell)
LMD+V model have been chosen as in the
papers~\cite{Nyffeler:2009tw,Jegerlehner:2009ry,Melnikov:2003xd}.  We
would like to stress again that our estimate of the $a_\mu^{{\rm
    LbyL}; \pi^0}$ uncertainty is given only by the propagation of the
errors of the newly fitted parameters listed in Table~\ref{tab:amu}
and therefore we may not reproduce the total uncertainties of
$a_\mu^{{\rm LbyL}; \pi^0}$ given in the original papers.

We can clearly see from Table~\ref{tab:amu} that for each given model
and each approach (JN or MV), there is a trend of reduction in the
error for $a_\mu^{{\rm LbyL}; \pi^0}$ (related only to the given model
parameters) by about half when going from A0 (PDG) to A1 (including
PrimEx) and by about another half when going from A1 to A2 (including
KLOE-2). This is mainly due to the improvement in the normalization of
the form factor (decay width $\pi^0 \to \gamma\gamma$), controlled by
the parameters $F_\pi$ or $\bar{h}_7$, respectively, but more data
also better constrain the other model parameters $M_V$ or $\bar{h}_5$,
respectively.

This trend of improvement is also visible in the last part of the
Table (LMD+V, $h_1 \neq 0$), when we fit the sets B0, B1 and B2 which
include the BaBar data. Furthermore, since we now have even more data
to fit, the final error on $a_\mu^{{\rm LbyL}; \pi^0}$ is improved
further, compared to the fits of LMD+V with $h_1 \neq 0$ of the data
sets A0, A1 and A2 only. This can be seen in the errors of $\bar{h}_5$
and $h_1$. On the other hand, the parameter $\bar{h}_7$, related to
the normalization of the form factor at the origin, is essentially
unchanged by the inclusion of the BaBar data at high $Q^2$. The
central values of the final results for $a_\mu^{{\rm LbyL}; \pi^0}$
are only slightly changed, if we include the BaBar data. They shift
only by about $-0.5 \times 10^{-11}$ compared to the corresponding
data sets A0, A1 and A2. This is due to a partial compensation in
$a_\mu^{{\rm LbyL}; \pi^0}$, when the central values for $\bar{h}_5$
and $h_1$ are changed, as already observed in
Ref.~\cite{Nyffeler:2009uw}.

Since the data sets A0, A1 and A2 without BaBar show the $1/Q^2$
fall-off, fitting $h_1$ as a free parameter in the LMD+V model leads
to a value compatible with zero, but with quite some large 
error. Similarly, the value of $\bar{h}_5$ is shifted a bit and its
error gets doubled compared to the fit of the data sets A0,A1 and A2
with the LMD+V model with $h_1=0$ in the second part of the table. The
final result for LbyL, however, only shifts by about $\pm 0.1 \times
10^{-11}$.

Finally, note that both VMD and LMD+V with $h_1 = 0$ can fit the data
sets A0, A1 and A2 for the transition form factor very well with
essentially the same $\chi^2$ per degree of freedom for a given data
set (see first and second part of the table).  Nevertheless, the
results for the pion-exchange contribution to hadronic LbyL scattering
differ by about $20~\%$ in these two models. For VMD the result is
about $a_\mu^{{\rm LbyL}; \pi^0} \sim 57.5 \times 10^{-11}$ and for
LMD+V with $h_1 = 0$ it is about $72.5 \times 10^{-11}$ with the JN
approach and about $80 \times 10^{-11}$ with the MV approach. This is
due to the different behavior, in these two models, of the fully
off-shell form factor ${\cal F}_{{\pi^0}^*\gamma^*\gamma^*}((q_1 +
q_2)^2, q_1^2, q_2^2)$ on all momentum variables, which enters for the
pion-exchange contribution in hadronic LbyL
scattering~\cite{Jegerlehner:2007xe,Jegerlehner:2008zza}.

We conclude that the KLOE-2 data with a total integrated luminosity of
$5$~fb$^{-1}$ will give a reasonable improvement in the part of the
$a_\mu^{{\rm LbyL}; \pi^0}$ error associated with the parameters
accessible via the $\Gamma_{\pi^0 \to \gamma\gamma}$ width and the
$\pi^0\gamma\gamma^\ast$ form factor $F(Q^2)$.  As stressed above,
depending on the modelling of the off-shellness of the pion, there
might be other, potentially larger sources of uncertainty which cannot
be improved by the KLOE-2 measurements.

\section{Conclusions}
\label{sec:conclusions}

A simulation of the KLOE-2  experiment with 1~year of data taking
was performed.
Numerical results indicate a feasibility of $\sim 1\%$ 
statistical error in the measurement of $\Gamma_{\pi^0 \to \gamma\gamma}$.
Such a precision is better than 
the current experimental world average and 
the theoretical accuracy.
The $\pi^0$ electromagnetic transition form factor 
$F(Q^2)$ in the region $0.01 < Q^2 < 0.1  \, \text{GeV}^{2}$
can be measured with a statistical error of $< 6\%$ in each bin. 
This low $Q^2$ measurement can test the consistency of the
models which have been fitted so far to the data from CELLO, CLEO and BaBar
at higher $Q^2$ and will serve as an important test of the strong
interaction dynamics at low energies.  
The proposed measurements with the KLOE-2 experiment can also have an
impact on the value and precision of the contribution of a neutral
pion exchange to the hadronic light-by-light scattering in the muon
$g-2$. 
We would like to stress that a realistic calculation of this
contribution is {\it not} the purpose of this letter.  The given
estimates for $a_\mu^{{\rm LbyL}; \pi^0}$ should only demonstrate,
within several approaches, an improvement of uncertainty, which will
be possible when the KLOE-2 data appear.

\section*{Appendix A: The KLOE-2 Collaboration}
\noindent
F. Archilli, D. Babusci, D. Badoni, I. Balwierz,
G. Bencivenni, C. Bini, C. Bloise, V. Bocci, F. Bossi,
P. Branchini, A. Budano, S. A. Bulychjev,
L. Caldeira~Balkest\aa hl, P. Campana, G. Capon,
F. Ceradini, P. Ciambrone, E. Czerwi\'nski, E. Dan\'e,
E. De Lucia, G. De Robertis, A. De Santis, G. De Zorzi,
A. Di Domenico, C. Di Donato, D. Domenici, O. Erriquez, G. Fanizzi, 
G. Felici, S. Fiore, P. Franzini, P. Gauzzi, G. Giardina, 
S. Giovannella, F. Gonnella, E. Graziani, F. Happacher, 
B. H\"oistad, L. Iafolla, E. Iarocci, M. Jacewicz, T. Johansson,
A. Kowalewska, V. Kulikov, A. Kupsc, J. Lee-Franzini,
F. Loddo, G. Mandaglio, M. Martemianov, M. Martini, M. Mascolo,
M. Matsyuk, R. Messi, S. Miscetti,
G. Morello, D. Moricciani, P. Moskal, F. Nguyen,
A. Passeri, V. Patera, I. Prado Longhi, A. Ranieri,
C. F. Redmer, P. Santangelo, I. Sarra, M. Schioppa,
B. Sciascia, A. Sciubba, M. Silarski, C. Taccini,
L. Tortora, G. Venanzoni, R. Versaci, W. Wi\'slicki,
M. Wolke, J. Zdebik

\subsection*{Acknowledgements}
\label{sec:Acknowl}
The authors wish to thank Catia Milardi from LNF Accelerator Division 
for the information on the DA$\Phi$NE optics and 
checking our results on the off-energy lepton tracking.
We profited from discussions with 
Simon Eidelman, Fred Jegerlehner, 
Wolfgang Kluge and Peter Lukin.
This work is a part of the activity of the ``Working Group on Radiative
Corrections and Monte
Carlo Generators for Low Energies'' \footnote{{\tt http://www.lnf.infn.it/wg/sighad/}}.
The partial support from 
the European Community-Research Infrastructure Integrating Activity 
``Study of Strongly Interacting Matter'' 
(acronym HadronPhysics2, Grant Agreement~n.~227431) under 
the Seventh Framework Program of EU,
and
Polish Ministry of Science and High Education
from budget for science for years 2010-2013: grant number N~N202~102638
is acknowledged.
This work was also partially supported by funding available from the Department
of Atomic Energy, Government of India, for the Regional Centre for
Accelerator - based Particle Physics (RECAPP), Harish-Chandra Research
Institute.
S.I. was also partially supported by the National Academy of Science
of Ukraine under contract $50/53 - 2011$.
A.N. and S.I. wish to thank INFN-LNF and  INFN-Roma ``Tor Vergata''
for the kind hospitality.
%


\end{document}